\documentclass[english]{sbrt}
\usepackage{url} 
\usepackage{array}
\usepackage{graphicx}
\usepackage[english]{babel}
\usepackage[utf8]{inputenc}
\usepackage{amsmath, amssymb}

\usepackage{cite}
\usepackage{xcolor}
\usepackage[acronyms,shortcuts]{glossaries}
\newcommand{\sumx}[2]{\sum\limits_{#1}^{#2}}
\newcommand{\bb}[1]{\mathbb{#1}}

\newcommand{\ma}[1]{\boldsymbol{#1}}
\definecolor{green}{rgb}{0.1,0.75,0.2}

\newacronym{2G}{2G}{second generation}
\newacronym{3G}{3G}{third generation}
\newacronym{4G}{4G}{fourth generation}
\newacronym{5G}{5G}{fifth generation}
\newacronym{B5G}{B5G}{beyond fifth generation}
\newacronym{6G}{6G}{sixth generation}
\newacronym{3GPP}{3GPP}{3$\text{rd}$~Generation Partnership Project}
\newacronym{LTE}{LTE}{long term evolution}
\newacronym{NR}{NR}{new radio}
\newacronym{LS}{LS}{least squares}

\newacronym{IRS}{IRS}{intelligent reconfigurable surface}
\newacronym{RIS}{RIS}{reconfigurable intelligent surface}
\newacronym{LIS}{LIS}{large intelligent surface}
\newacronym{SDS}{SDS}{software-defined surface}

\newacronym{D2D}{D2D}{device-to-device}
\newacronym{BS}{BS}{base station}
\newacronym{UE}{UE}{user equipment}

\newacronym{SU}{SU}{single-user}
\newacronym{MU}{MU}{multi-user}
\newacronym{SISO}{SISO}{single-input single-output}
\newacronym{MISO}{MISO}{multiple-input single-output}
\newacronym{SIMO}{SIMO}{single-input multiple-output}
\newacronym{MIMO}{MIMO}{multiple-input multiple-output}

\newacronym{CSI}{CSI}{channel state information}
\newacronym{LOS}{LOS}{line-of-sight}
\newacronym{NLOS}{NLOS}{non-line-of-sight}

\newacronym{QoS}{QoS}{quality-of-service}
\newacronym{SE}{SE}{spectral efficiency}
\newacronym{EE}{EE}{energy efficiency}
\newacronym{SINR}{SINR}{signal to interference plus noise ratio}
\newacronym{SNR}{SNR}{signal to noise ratio}

\newacronym{ProSe}{ProSe}{proximity services}
\newacronym{NSPS}{NSPS}{national security and public safety}

\newacronym{RRM}{RRM}{radio resource management}
\newacronym{MS}{MS}{mode selection}
\newacronym{RA}{RA}{resource allocation}
\newacronym{PC}{PC}{power control}

\newacronym{BCD}{BCD}{block coordinate descent}

\newacronym{RF}{RF}{radio frequency}
\newacronym{AWGN}{AWGN}{additive white Gaussian noise}
\newacronym{MRC}{MRC}{maximum ratio combining}

\newacronym{AF}{AF}{amplify-and-forward}
\newacronym{DF}{DF}{decode-and-forward}
\newacronym{DFT}{DFT}{discrete Fourier transform}
\newacronym{TX}{TX}{transmitter}
\newacronym{RX}{RX}{receiver}
\newacronym{ALS}{ALS}{alternating least squares}
\newacronym{BALS}{BALS}{bilinear alternating least squares}
\newacronym{SVD}{SVD}{singular value decomposition}
\newacronym{HOSVD}{HOSVD}{high order singular value decomposition}
\newacronym{THOSVD}{THOSVD}{truncated high order singular value decomposition}
\newacronym{PARAFAC}{PARAFAC}{PARAllel FACtors}
\newacronym{AOD}{AOD}{angle of departure}
\newacronym{AOA}{AOA}{angle of arrival}
\newacronym{URA}{URA}{uniform rectangular array} 
\newacronym{ADR}{ADR}{achievable data rate}
\newacronym{NMSE}{NMSE}{normalized mean square error}
\newacronym{SER}{SER}{symbol error rate}
\newacronym{LRA}{LRA}{low-rank approximation}

\newacronym{ULA}{ULA}{uniform linear array}
\newacronym{mmWave}{mmWave}{milimiter-wave}
\newacronym{CS}{CS}{compressed sensing}
\newacronym{OFDM}{OFDM}{orthogonal frequency division multiplexing}
\newacronym{PIN}{PIN}{positive-intrinsic-negative}
\newacronym{BD-RIS}{BD-RIS}{beyond diagonal reconfigurable intelligent surface}
\newacronym{LS-Kron}{LS-Kron}{least squares Kronecker factorization}

\newacronym{BTALS}{BTALS}{block Tucker alternating least squares}
\newacronym{BTKF}{BTKF}{block Tucker Kronecker factorization}
\newacronym{PALS}{PALS}{PARAFAC alternating least squares}
\newacronym{PKF}{PKF}{PARAFAC Khatri-Rao factorization}
\newacronym{ISAC}{ISAC}{integrated sensing and communication}
\renewcommand{\baselinestretch}{0.9}
\begin{document}

\title{Performance Evaluation of Beyond Diagonal RIS under Hardware Impairments}

\author{Jose Carlos da Silva Filho, Josué V. de Araújo, Bruno Sokal, André L. F. de Almeida 
\thanks{Jose Carlos da Silva Filho, Josué V. de Araújo, Bruno Sokal, André L. F. de Almeida are with Grupo de Pesquisa em Telecomunicações Sem Fio (GTEL), Departamento de Engenharia de Teleinformática, Universidade Federal do Ceará, Fortaleza-CE. E-mails: jcarlos.sf237@gmail.com, josue.vas@alu.ufc.br, brunosokal@gtel.ufc.br, andre@gtel.ufc.br. The authors thank the partial support of FUNCAP under grant ITR-0214-00041.01.00/23, and the National Institute of Science and Technology (INCT-Signals) sponsored by Brazil’s
National Council for Scientific and Technological Development (CNPq) under grant 406517/2022-3. This work is also supported by CNPq under grant 312491/2020-4 and by Ericsson Research, Sweden, and Ericsson Innovation Center, Brazil, under UFC.52 Technical Cooperation Contract Ericsson/UFC.}%
}

\maketitle

\markboth{XLIII BRAZILIAN SYMPOSIUM ON TELECOMMUNICATIONS AND SIGNAL PROCESSING - SBrT 2025, SEPTEMBER 29TH TO OCTOBER 2ND, NATAL, RN}{}

\begin{abstract}
\Ac{BD-RIS} improves the traditional \ac{RIS} architecture functionality by interconnecting elements for advanced wave control. However, real‐world implementations face hardware imperfections, such as impedance mismatches and varactor nonidealities, which can degrade overall system performance. In this paper, we propose three hardware impairment models that directly affect the BD-RIS scattering matrix structure and evaluate their impact on the channel estimation accuracy using the \ac{NMSE} as a performance metric. The proposed impairment models consider imperfections affecting self-impedances, mutual impedances, or both. Our results reveal how each impairment type degrades the system performance, allowing us to identify scenarios where the traditional RIS can outperform the BD‐RIS.

\end{abstract}

\section{Introduction}
In a typical communication system, the transmitted signal suffers from several impairments due to the surrounding environment, including attenuation and scattering, which is generally caused by multipath propagation, one of the main limiting factors in this wireless communication system. An emerging technology capable of mitigating these effects and improving propagation conditions is \ac{RIS} which is a man-made surface composed of electromagnetic materials that consists of a 2D array with many low-cost passive elements that can control the electromagnetic properties of radio-frequency waves, such as phase, amplitude, and frequency \cite{basar_wireless_ris,gil_parafac,tensor_irs}. In general, the propagation of electromagnetic waves in wireless communication channels is not controlled between the transmission and reception processes. However, the RIS can "reprogram" the channel during this stage. Classical RIS architectures consider each element independent of the others, i.e., no interconnection among elements exists. This design translates into the signal model, where the scattering matrix has a diagonal structure. As a result, some limitations arise, such as limited beamforming capabilities and restricted degrees of freedom to control the direction of signal reflection\cite{survey_irs}.

Recent studies have introduced beyond diagonal reconfigurable intelligent surface (BD-RIS) to overcome conventional RIS's limitations and enhance system performance \cite{bd_ris_magazine}. In BD-RIS architectures, the reflecting elements are intentionally interconnected, leading to a non-diagonal signal model matrix. More precisely, in the  BD-RIS architecture, the connections are primarily defined as \cite{shen_analysis,bd_ris_architecture,li_dynamic_bd_ris,bd_ris_group,almeida_sokal_bd_ris_ce_TSP_2025,li_channel_estimation} 1) group-connected, where only a portion of the RIS elements are interconnected, leading to the signal model a block-diagonal matrix as scattering matrix, and 2) fully-connected, which represents the design that all RIS elements are connected. In this case, the signal model scattering matrix is complete, i.e., with all non-zero entries. This non-diagonal design of the scattering matrix provides additional degrees of freedom and enables a more flexible control over the signal properties. This includes joint manipulation of phase and amplitude and the ability to create more reflection patterns, which, in specific scenarios, allow BD-RIS to outperform the conventional diagonal RIS.

Despite the promising theoretical performance of BD-RIS, most existing studies are still based on idealized models with perfect operating conditions. In many cases, however, such models do not fully capture the practical imperfections possible in real-world implementations, such as impedance mismatches and losses in the passive interconnected components of the internal circuits of the BD-RIS structure. These limitations highlight the need for more realistic performance evaluations considering physical/electronic impairments.

In this paper, we evaluate the performance of BD-RIS-assisted communication systems in the presence of circuit-based hardware impairments. We start from the baseline signal model of \cite{bd_ris_group} by including the hardware impairment models and assess the solution of \cite{bd_ris_group} under imperfections. We consider three hardware impairment models directly affecting the self-impedances, mutual impedances, or both. Simulation results show how these imperfections affect the channel estimation performance when state-of-the-art algorithms are used.

\section{Notation and properties}
To facilitate the understanding of the material presented throughout the paper, we define the key mathematical notations and properties used. Scalars are denoted by non-bold lowercase letters, such as \(a\), column vectors are denoted by bold lowercase letters, such as \(\boldsymbol{a}\), and matrices are represented by bold
uppercase letters, such as \(\boldsymbol{A}\). \(\boldsymbol{I}_{K}\) represents an identity matrix of size $K \times K$, while $\ma{1}_{N \times N}$ stands for a all-ones matrix of size $N \times N$. The superscripts \(\{\cdot\}^{\mathrm{T}},\{\cdot\}^{*}\) and \(\{\cdot\}^{\mathrm{H}}\) denote the transpose, conjugate, and conjugate transpose, respectively. The operator \(\|\cdot\|_{F}\) refers to the Frobenius norm of a matrix or tensor, and \(\mathbb{E}\{\cdot\}\) denotes the expectation operator. The operator \(\operatorname{diag}(\boldsymbol{a})\) converts a vector \(\boldsymbol{a}\) into a diagonal matrix. From a set of \(Q\) matrices \(\boldsymbol{X}^{(q)} \in \mathbb{C}^{M \times N}\), with \(q=\{1, \ldots, Q\}\), we can construct a block diagonal matrix as \(\boldsymbol{X}=\operatorname{blkdiag}\left(\boldsymbol{X}^{(1)}, \ldots, \boldsymbol{X}^{(Q)}\right) \in \mathbb{C}^{M Q \times N Q}\). Moreover, the operator \(\operatorname{vec}(\boldsymbol{A})\) converts a matrix \(\boldsymbol{A} \in \mathbb{C}^{M \times N}\) to a column vector \(\boldsymbol{a} \in \mathbb{C}^{M N \times 1}\) by stacking its columns on top of each other, while the operator unvec\((\boldsymbol{a})_{M \times N}\) reverses this operation, restoring the matrix \(\boldsymbol{A} \in \mathbb{C}^{M \times N}\). The symbols \(\circ\), \(\odot\), \(\otimes\), and \(\diamond\) denote the outer product, the Hadamard product, the Kronecker product, and the Khatri-Rao product (also known as the column-wise Kronecker product), respectively. Some useful properties that will be exploited in this paper are the following.
\begin{align}
\operatorname{vec}(\boldsymbol{A} \boldsymbol{B} \boldsymbol{C}) & =\left(\boldsymbol{C}^{\mathrm{T}} \otimes \boldsymbol{A}\right) \operatorname{vec}(\boldsymbol{B}),\label{eq:vec_ABC} \\
\operatorname{vec}(\boldsymbol{A} \otimes \boldsymbol{B}) & =\boldsymbol{P}(\operatorname{vec}(\boldsymbol{A}) \otimes \operatorname{vec}(\boldsymbol{B})), \\
\operatorname{vec}\left(\boldsymbol{a} \boldsymbol{b}^{\mathrm{T}}\right) & =\boldsymbol{b} \otimes \boldsymbol{a},
\end{align}
where the vectors and matrices involved have compatible dimensions in each case, and \(\boldsymbol{P}\) is a permutation matrix. The Khatri-Rao product of the matrices \(\boldsymbol{X} \in \mathbb{C}^{I \times R}\) and \(\boldsymbol{Y} \in \mathbb{C}^{J \times R}\), is defined as
\begin{align}
\boldsymbol{Z}=\boldsymbol{X} \diamond \boldsymbol{Y}=\left[\boldsymbol{x}_{1} \otimes \boldsymbol{y}_{1}, \ldots, \boldsymbol{x}_{R} \otimes \boldsymbol{y}_{R}\right] \in \mathbb{C}^{J I \times R},
\label{kr}
\end{align}
where \(\boldsymbol{x}_{r}\) and \(\boldsymbol{y}_{r}\), are the \(r\)-th column of \(\boldsymbol{X}\) and \(\boldsymbol{Y}\), respectively, for \(r=\{1, \ldots, R\}\). 

%%%%%%%%%%%%%%%%%%%%%%%%%%%%%%%%%%%%%%%%%%%%%%%%%%%%%%%%%%%%%%%%%%%%%%%%%%%%%%%%%%%%%%%%%%%%%%%%%%%%%%%%%%%%%%%%%%%%%%%%%%%%%%%%%%%%%%%%%%%%%%%%%%%%%%%%%%%%%%%%%%%%%%%%%%%%%%%%%%%%%%%%%%%%%%%%%%%%%%%%%%%%%%%%%%%%%%%%%%%%%%%%%%%%%%%%%%%%%%%%%%%%%%%%%%%%%%%%%%%%%%%%%%%%%%%%%%%%%%%%%%%%%%%%%%%%%%%%%%%%%%%%%%%%%%%%%%%%%%%%%%%%%%%%%%%%%%%%%%%%%%%%%%%%%%%%%%%%%%%%%%%%%%%%%%%%%%%%%%%%%%%%%%%%%%%%%%%%%%%%%%%%%%%%%%%%%%%%%%%%%%%%%%%%%%%%%%%%%%%%%%%%%%%%%%%%%%%%%%%%%%%%%%%%%%%%%%%%%%%%%%%%%%%%%%%%%%%%%%%%%%%%%%%%%%%%%%%%%%%%%%%%%%%%%%%%%%%%%%%%%%%%%%%%%%%%%%%%%%%%%%%%%%%%%%%%%%%%%%%%%%%%%%%%%%%%%%%%%%%%%%%%%%%%%%%%%%%%%%%%%%%%%%%%%%%%%%%%%%%%%%%%%%%%%%%%%%%%%%%%%%%%%%%%%%%%%%%%%%%%%%%%%%%%%%%%%%%%%%%%%%%%%%%

\section{System Model}
%%\in
We consider the system model described  in \cite{decoupled_bd_ris}, i.e., a \ac{MIMO} communication system assisted by a BD-RIS composed of $N$ elements, divided into $Q$ groups, each of size $\bar{N}$, i.e., $N = \bar{N}Q$. The transmitter (Tx) and the receiver (Rx) have $M_T$ and $M_R$ antennas, respectively. The direct link (TX-RX) is assumed to be blocked for convenience. The transmission duration is over $T$ time-slots, from which the transmitted pilots are defined as $\ma{X} = [\ma{x}_1,\ldots,\ma{x}_T]  \in \bb{C}^{M_T \times T}$.  The received signal on the $t$-th transmitted time slot is given by

\begin{align}
    \ma{y}_t = \sumx{q=1}{Q} \ma{G}^{(q)}\ma{S}_t^{(q)}\ma{H}^{(q)\text{T}}\ma{x}_t + \ma{b}_t \in \bb{C}^{M_R \times 1},
\end{align}
where, $\ma{H}^{(q)} \in \bb{C}^{M_T \times \bar{N}}$ and $\ma{G}^{(q)} \in \bb{C}^{M_R \times \bar{N}}$ are the $q$-th block matrix of the TX-RIS and RIS-RX channels, i.e., $\ma{H} = [\ma{H}^{(1)},\ldots,\ma{H}^{(Q)}] \in \bb{C}^{M_T \times \bar{N}Q}$ and $\ma{G} = [\ma{G}^{(1)},\ldots,\ma{G}^{(Q)}] \in \bb{C}^{M_R \times \bar{N}Q}$.  $\ma{S}_t^{(q)} \in \bb{C}^{\bar{N} \times \bar{N}}$ represents the BD-RIS scattering matrix of the $q$-th group at the $t$-th time slot.     

% We consider that the channel remains constant throughout the entire length-T pilot sequence. The matrices holding the channel coefficients associated with the Tx-RIS and RIS-Rx links are defined as \(\boldsymbol{H} \in \mathbb{C}^{M_{T} \times N}\) and \(\boldsymbol{G} \in \mathbb{C}^{M_{R} \times N}\), respectively.  

% Let \(\boldsymbol{S}_{t} \in \mathbb{C}^{N \times N}\) be the unitary scattering matrix of the BD-RIS at time $t$, which means that \(\boldsymbol{S}_{t}^{\mathrm{H}} \boldsymbol{S}_{t}=\boldsymbol{I}_{N}\)\cite{bd_ris_group}, and the additive white Gaussian noise (AWGN) is represented by \(\boldsymbol{b}_{t}\).

To model a possible real-world hardware imperfection affecting the impedances, we consider different types of impairments that are given by a matrix $\mathbf{E} \in \mathbb{C}^{N \times N}$ in which is detailed in Section \ref{Sec:Impairments}. For the sake of convenience, we consider that hardware impairment is constant over the transmission of the $T$ time slots. Thus, the real observed BD-RIS scattering matrix, at the $t$-th time slot, can be denoted as
\begin{align}
\label{eq:impaired_S}
    \bar{\ma{S}}_t = \ma{S}_{t} \odot \ma{E} \in \bb{C}^{N \times N},
\end{align}
where $\ma{S}_t$ is the ideal BD-RIS scattering matrix at the $t$-th time slot, for $t = 1,\ldots,T$.
Hence, the signal received in the $t$-th time slot can be expressed as
\begin{align}
\label{eq:modelo_geral_with_error}
\boldsymbol{y}_t &=  \sumx{q=1}{Q} \ma{G}^{(q)}\left(\ma{S}_t^{(q)} \odot \ma{E}\right)\ma{H}^{(q)\text{T}}\ma{x}_t + \ma{b}_t \in \bb{C}^{M_R \times 1}, \\
&= \sumx{q=1}{Q} \ma{G}^{(q)}\bar{\ma{S}}_t\ma{H}^{(q)\text{T}}\ma{x}_t + \ma{b}_t. 
\end{align}

% Assuming group-connected architectures, we uniformly divide the $N$ elements into $Q$ groups, where \(\bar{N} = \frac{N}{Q}\) denotes the number of elements connected \cite{bd_ris_group}. In addition, in this case, the scattering and impairment matrices have a block diagonal structure 
% \[\boldsymbol{S}_{t}=\operatorname{blkdiag}\left(\boldsymbol{S}_{t}^{(1)}, \ldots \boldsymbol{S}_{t}^{(Q)}\right) \in \mathbb{C}^{N \times N},\] with every \(q\)-th scattering matrix 
% \(\boldsymbol{S}_{t}^{(q)}\) satisfying \(\boldsymbol{S}_{t}^{(q) \mathrm{H}} \boldsymbol{S}_{t}^{(q)}=\boldsymbol{I}_{\bar{N}}\), and \[\mathbf{E} = \mathrm{blkdiag} \left( \mathbf{E}^{(1)}, \ldots, \mathbf{E}^{(Q)} \right)\in \mathbb{C}^{N \times N}.\]

% Similarly, we also interpret the channel matrices \(\boldsymbol{H}\) and \(\boldsymbol{G}\) as a concatenation of $Q$ group submatrices defined as
% \[
% \begin{aligned}
% \boldsymbol{H}^{(q)} =[\mathbf{H}]_{:,(q-1) \bar{N}+1: q \bar{N}}\in \mathbb{C}^{M_{T} \times \bar{N}}, q=1, \ldots, Q~ \\
% \boldsymbol{G}^{(q)}=[\mathbf{G}]_{:,(q-1) \bar{N}+1: q \bar{N}} \in \mathbb{C}^{M_{R} \times \bar{N}}, q=1, \ldots, Q,
% \end{aligned}
% \]
% which allows us to rewrite the $t$-th received signal as
\begin{figure}[!t]
    \centering
    \includegraphics[width=0.45\textwidth]{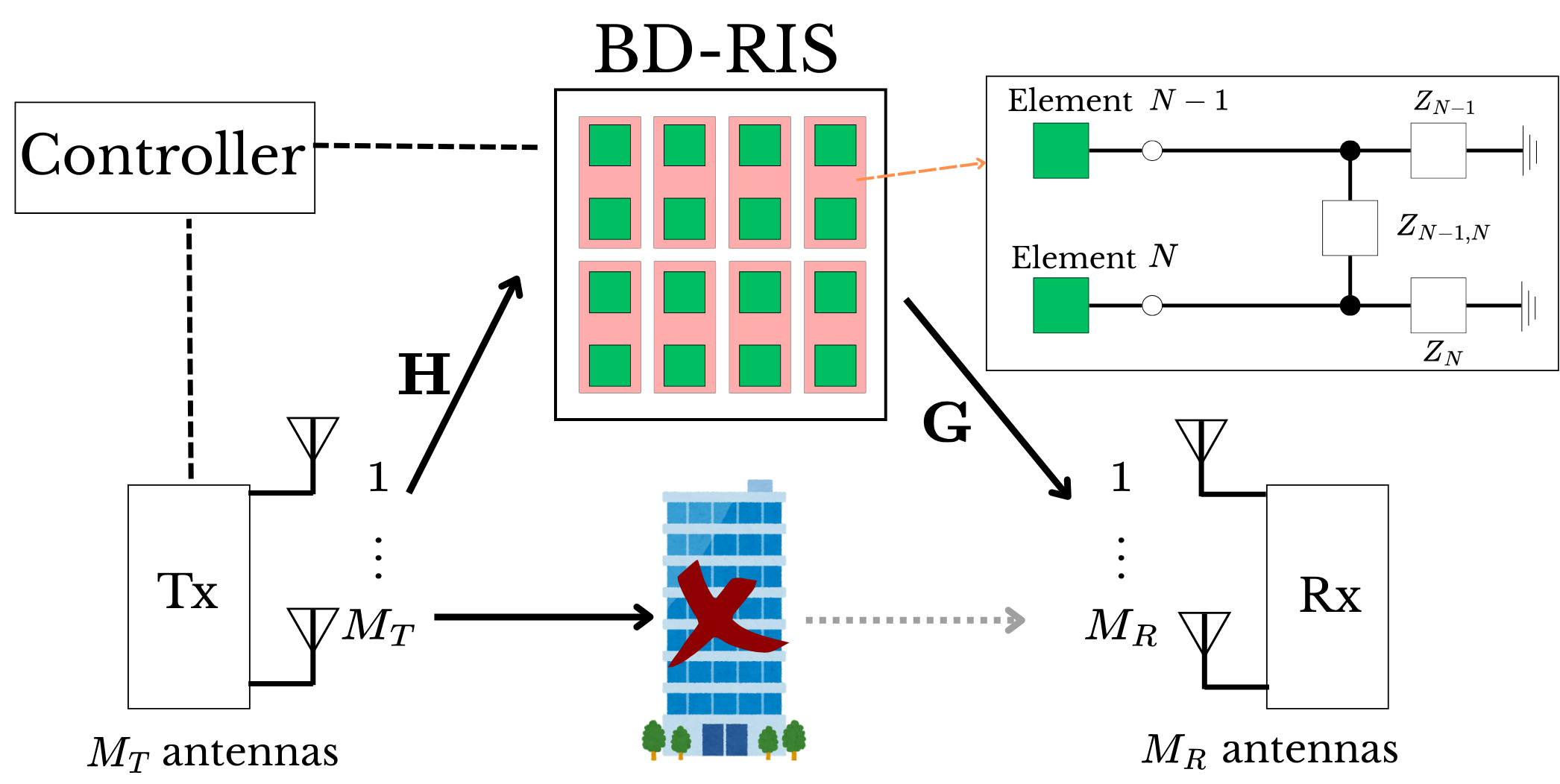}
    \caption{System model of a MIMO communication aided by a BD-RIS.}
    \label{fig:system_model}
\end{figure}
\begin{figure*}[!h]
    \centering
    \includegraphics[width=.75\textwidth]{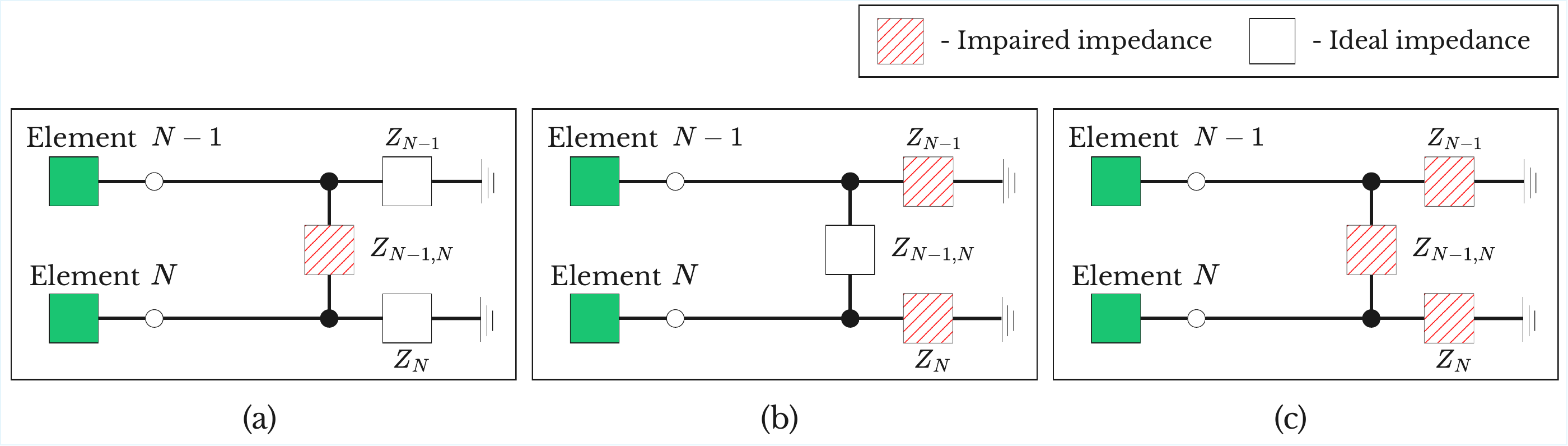}
    \caption{Illustration of the three impairment types: (a) Type 1 – only the impedances that connect the elements are under impairment; (b) Type 2 – the elements impedances are under impairments, but the connections are ideal; (c) Type 3 – all the impedances are under impairments.}
    \label{fig:impairments}
\end{figure*}
% \begin{equation}\label{eq:modelo_gera_groupsl}
% \boldsymbol{y}_t = \sum_{q=1}^{Q} \boldsymbol{G}^{(q)} \bar{\mathbf{S}}^{(q)}_t \boldsymbol{H}^{(q) \mathrm{T}} \boldsymbol{x}_{t} + \boldsymbol{b}_{t},
% \end{equation}
% where \(\boldsymbol{\bar{S}}_{t}^{(q)}\) is the \(q\)-th group of the scattering matrix affected by the impairment, for \(q=\{1, \ldots, Q\}\).
Using the property \eqref{eq:vec_ABC}, as demonstrated in \cite{decoupled_bd_ris}, the noise free received signal can be written as
\begin{align}
\boldsymbol{y}_{t}
&= \sum_{q=1}^{Q} \big( \operatorname{vec} \big( \bar{\boldsymbol{S}}_{t}^{(q)} \big)^{\mathrm{T}} \otimes \boldsymbol{x}_{t}^{\mathrm{T}} \otimes \boldsymbol{I}_{M_{R}} \big) \boldsymbol{c}^{(q)},
\end{align}
% \begin{align}
% \boldsymbol{y}_{t}^{(k)} &= \sum_{q=1}^{Q} \big( \boldsymbol{x}_{t}^{\mathrm{T}} \otimes \boldsymbol{I}_{M_{R}} \big) \operatorname{vec} \big( \boldsymbol{G}^{(q)} \bar{\boldsymbol{S}}_{t}^{(k,q)} \boldsymbol{H}^{(q)\mathrm{T}} \big) \\
% &= \sum_{q=1}^{Q} \big( \boldsymbol{x}_{t}^{\mathrm{T}} \otimes \boldsymbol{I}_{M_{R}} \big) \big( \boldsymbol{H}^{(q)} \otimes \boldsymbol{G}^{(q)} \big) \operatorname{vec} \big( \bar{\boldsymbol{S}}_{t}^{(k,q)} \big) \\
% &= \sum_{q=1}^{Q} \big( \operatorname{vec} \big( \bar{\boldsymbol{S}}_{t}^{(k,q)} \big)^{\mathrm{T}} \otimes \boldsymbol{x}_{t}^{\mathrm{T}} \otimes \boldsymbol{I}_{M_{R}} \big) \boldsymbol{c}^{(q)} 
% \end{align}
where $\boldsymbol{c}^{(q)} = \operatorname{vec} \left( \boldsymbol{H}^{(q)} \otimes \boldsymbol{G}^{(q)} \right) \in \mathbb{C}^{M_{R} M_{T} \bar{N}^{2} \times 1}$. Applying property \eqref{kr}, collecting the $T$ time slots in a single observation vector, and removing the summation, we obtain the following equivalent expression 
\begin{equation}
\begin{aligned}
\boldsymbol{y} &= \left[\boldsymbol{y}_{1}^{\mathrm{T}}, \ldots, \boldsymbol{y}_{T}^{\mathrm{T}} \right]^{\mathrm{T}} = \left[ \left(  \bar{\ma{S}}^{\prime} \diamond \boldsymbol{X} \right)^{\mathrm{T}} \otimes \boldsymbol{I}_{M_{R}} \right] \boldsymbol{c} \\
&= \left[ \boldsymbol{\bar{\Omega}} \otimes \boldsymbol{I}_{M_{R}} \right] \boldsymbol{c} \in \mathbb{C}^{M_{R}T \times 1},
\end{aligned}
\end{equation}
where $\boldsymbol{\bar{\Omega}} = \left( \bar{\ma{S}}^{\prime} \diamond \boldsymbol{X} \right)^{\mathrm{T}} \in \mathbb{C}^{T \times M_T \bar{N}^2 Q}$ is the combined BD-RIS impaired scattering matrix and the transmitted pilot. The matrix $\bar{\ma{S}}^{\prime} = \left[ \bar{\boldsymbol{s}}_1, \ldots, \bar{\boldsymbol{s}}_T \right] \in \mathbb{C}^{\bar{N}^2 Q \times T}$ is built from the vectorized, group-wise impaired scattering matrices, i.e., $\ma{s}_t = [\text{vec}(\bar{\ma{S}}^{(1)}_t)^{\text{T}},\ldots,\text{vec}(\bar{\ma{S}}^{(Q)}_t)^{\text{T}}]^{\text{T}} \in \bb{C}^{\bar{N}^2Q \times 1}$ for $t = 1\ldots,T$. The combined channel vector is defined as $\boldsymbol{c} = \left[ \boldsymbol{c}^{(1)\mathrm{T}}, \ldots, \boldsymbol{c}^{(Q)\mathrm{T}} \right]^{\mathrm{T}} \in \mathbb{C}^{M_R M_T \bar{N}^2 Q \times 1}$, for $q = 1, \ldots, Q$. Then the authors in \cite{decoupled_bd_ris} formulates the following  linear \ac{LS} estimation problem
\[
\hat{\boldsymbol{c}} = \underset{\boldsymbol{c}}{\operatorname{argmin}} \left\| \boldsymbol{y} - \left( \boldsymbol{\Omega} \otimes \boldsymbol{I}_{M_R} \right) \boldsymbol{c} \right\|^2
,\]
where $\ma{\Omega} = (\ma{S} \diamond \ma{X})^{\text{T}}$ its the ideal combined BD-RIS scattering matrix $\ma{S}$ with the pilot matrix $\ma{X}$. Since $\ma{\Omega}$ is known at the receiver, it can be optimally designed, from which $\boldsymbol{\Omega}^{\text{H}}\boldsymbol{\Omega} = \ma{I}_{M_T\bar{N}^2Q}$. As solution for the \ac{LS} problem, the authors in \cite{decoupled_bd_ris} derived a matched filter, i.e.,
\begin{align}
    \label{eq:sol_LS} \hat{\ma{c}} = \frac{\bar{N}}{T} (\ma{\Omega} \otimes \ma{I}_{M_R})^{\text{H}}\ma{y}.
\end{align}
The resulting estimate aggregates all the $Q$ BD-RIS groups and can be approximately expressed as
\begin{align}
    \label{eq:c_sep}  \hat{\ma{c}} \approx [\text{vec}(\ma{H}^{(1)} \otimes \ma{G}^{(1)})^{\text{T}},\ldots,\text{vec}(\ma{H}^{(Q)} \otimes \ma{G}^{(Q)})^{\text{T}}]^{\text{T}}.
\end{align}
% \[
% \hat{\boldsymbol{c}} \approx \begin{bmatrix}
% \operatorname{vec}\big( \boldsymbol{H}^{(1)} \otimes \boldsymbol{G}^{(1)} \big)^{\mathrm{T}} &
% \cdots &
% \operatorname{vec}\big( \boldsymbol{H}^{(Q)} \otimes \boldsymbol{G}^{(Q)} \big)^{\mathrm{T}}
% \end{bmatrix}^{\mathrm{T}}.
% \]
To recover a unique solution for $\boldsymbol{c}$, the number of pilot symbols must satisfy $T \geq M_T \bar{N}^2 Q$ \cite{decoupled_bd_ris}. Also, the authors of \cite{decoupled_bd_ris} proposed a decoupling method that exploits the structure of (\ref{eq:c_sep}). However, the solution discussed in \cite{decoupled_bd_ris} considers the ideal scattering matrix $\ma{S}$, thus the matched filter in (\ref{eq:sol_LS}) is efficient. However, the scattering matrix may be non-ideal, containing impairments, as shown in (\ref{eq:impaired_S}). Thus, the solution in (\ref{eq:sol_LS}) translates to
\begin{align}
    \label{eq:real_filter} \hat{\ma{c}} = \frac{\bar{N}}{T} (\ma{\Omega} \otimes \ma{I}_{M_R})^{\text{H}} \left(( \boldsymbol{\bar{\Omega}} \otimes \boldsymbol{I}_{M_{R}} ) \boldsymbol{c} + \ma{b}\right),
\end{align}
where $(\ma{\Omega} \otimes \ma{I}_{M_R})^{\text{H}}( \boldsymbol{\bar{\Omega}} \otimes \boldsymbol{I}_{M_{R}} )= \ma{\Omega} \boldsymbol{\bar{\Omega}}^{\text{H}} \otimes \ma{I}_{M_R}  \neq \ma{I}_{M_RM_T\bar{N}^2Q}$. In this context, our goal is to evaluate the accuracy of channel estimation for BD-RIS by taking into account the mismatch between the ideal scattering matrix $\ma{S}$ and the actual imperfect scattering matrix $\bar{\ma{S}}$. To our knowledge, this is the first paper that studies the impact of these hardware imperfection models on the channel estimation performance for BD-RIS-assisted communication systems. In the following, we describe three hardware impairment models and their implications on the signal model.

\section{BD-RIS hardware impairments}\label{Sec:Impairments}

In this section, we discuss the possible sources of the impairments that can affect RIS, especially the \ac{BD-RIS}. Later, we discuss the impact of those impairments on the signal model in three different ways. We summarize the source of BD-RIS impairments as:
\begin{itemize}
\item \textbf{Impedance mismatch}: Variations in the physical or electrical properties of the RIS elements can lead to unintended impedance values, affecting individual self-impedances and/or mutual impedances.
    
\item \textbf{Defective varactors}: RIS structures are often made of varactor diodes\cite{varactor_model},\cite{paulo_impairments} to adjust the phase and amplitude of the reflected signal. Fabrication defects or control inaccuracies may distort the desired nominal response.
    
\item \textbf{Thermal and environmental effects}: Temperature fluctuations, humidity, and mechanical stress can alter the electrical characteristics of the circuit components over time, leading to dynamic impairments.
\end{itemize}

These practical effects highlight the importance of incorporating realistic models, such as \cite{imperfection_paulo}, into analyzing BD-RIS-assisted communication systems, enabling a more accurate performance evaluation and designing more robust solutions. To evaluate the actual performance of the proposed system, we consider the presence of three different impairment models that can affect the scattering matrix of the BD-RIS. In all cases, the impairment (error) matrix is represented by the matrix $\mathbf{E} \in \mathbb{C}^{N \times N}$. Then, the real scattering matrix, on the time slot $t$, is described by Eq. (\ref{eq:impaired_S}). This model allows us to individually select which scattering matrix elements have their amplitude and phase distorted. Note that if a given element is not affected, the corresponding entry in \(\mathbf{[E}_t]_{i,j}\) is set to $1$.
% , otherwise, it is modeled as a complex number that reflects possible amplitude and phase distortions.

Given the group-connected architecture of the BD-RIS, the global impairment matrix is defined as
\[\mathbf{E} = \mathrm{blkdiag} \left( \mathbf{E}^{(1)}, \ldots, \mathbf{E}^{(Q)} \right)\in \mathbb{C}^{N \times N},\]
where each block \(\mathbf{E}^{(q)} \in \mathbb{C}^{\bar{N} \times \bar{N}}\) represents the impairment affecting group $q$ of the BD-RIS. To maintain a physical reciprocity of energy, all impairment groups preserve the Hermitian symmetry of the system when applicable, which means that \(
\mathbf{E}^{(q)} = (\mathbf{E}^{(q)})^{\mathrm{H}}
\), i.e., \(
[\mathbf{E}^{(q)}]_{i,j} = [\mathbf{E}^{(q)}]_{j,i}^*
\) for $q = 1,\ldots,Q$. 
We consider three types of hardware impairments that may affect the interconnections of the BD-RIS, as follows.

\subsection{Type 1}
The first impairment type corresponds to assuming that only the mutual impedances interconnecting the BD-RIS elements are affected. This corresponds to assuming the off-diagonal elements of the impairment matrix $\ma{E}$ are composed of random error terms while the main diagonal ones are equal to 1. In other words, the Type $1$ impairment affects only the off-diagonal elements of the scattering matrix $\ma{S}$. Mathematically, the Type $1$ impairment matrix $\ma{E}$ is constructed as follows
\[
[\mathbf{E}^{(q)}]_{i,j} =
\begin{cases}
\alpha_{ij} e^{j\phi_{ij}}, & \text{if } i \neq j\\
1, & \text{otherwise},
\end{cases}
\]
where \(\alpha_{ij} \in (0, 1]\) and \(\phi_{ij}\in [0, 2\pi)\) represent the amplitude and phase distortions, respectively, and for $q = 1,\ldots, Q$. For the Type $1$ impairment, the maximum number of affected impedances is $\frac{N(\bar{N}-1)}{2} = \frac{\bar{N}^2Q - N}{2}$, corresponding to the off-diagonal entries.

\subsection{Type 2}
In this case, we assume that only the self-impedances of the BD-RIS elements are affected. This corresponds to assuming the main diagonal elements of the impairment matrix $\ma{E}$ are composed of random error terms while the off-diagonal ones are equal to 1. Hence, the Type $2$ impairment affects only the main diagonal elements of the scattering matrix $\ma{S}$. The  matrix $\ma{E}$ for the Type $2$ impairment is then given as
\[
[\mathbf{E}^{(q)}]_{i,j} =
\begin{cases}
\alpha_{ij} e^{j\phi_{ij}}, & \text{if } i = j \\
1, & \text{otherwise}.
\end{cases}
\] for $q = 1,\ldots, Q$. In this case, up to \(N\) impedances can be affected, corresponding to the diagonal of the matrix.

\subsection{Type 3}
In this impairment type, we assume all the impedances (self and mutual) connecting the BD-RIS elements are affected, which corresponds to a fully random impairment matrix $\ma{E}$, affecting all the scattering elements. In this case, we have
\[
[\mathbf{E}^{(q)}]_{i,j} = \alpha_{ij} e^{j\phi_{ij}}, \,\,\forall \, i,j = 1,\ldots, \bar{N}, \,\text{and} \, \, q = 1,\ldots, Q.
\]
The maximum number of potentially affected impedances in this case is $N+\frac{N(\bar{N}-1)}{2}$.

\section{Simulation Results}
\begin{figure}[!t]
    \centering
    \includegraphics[width=0.41\textwidth]{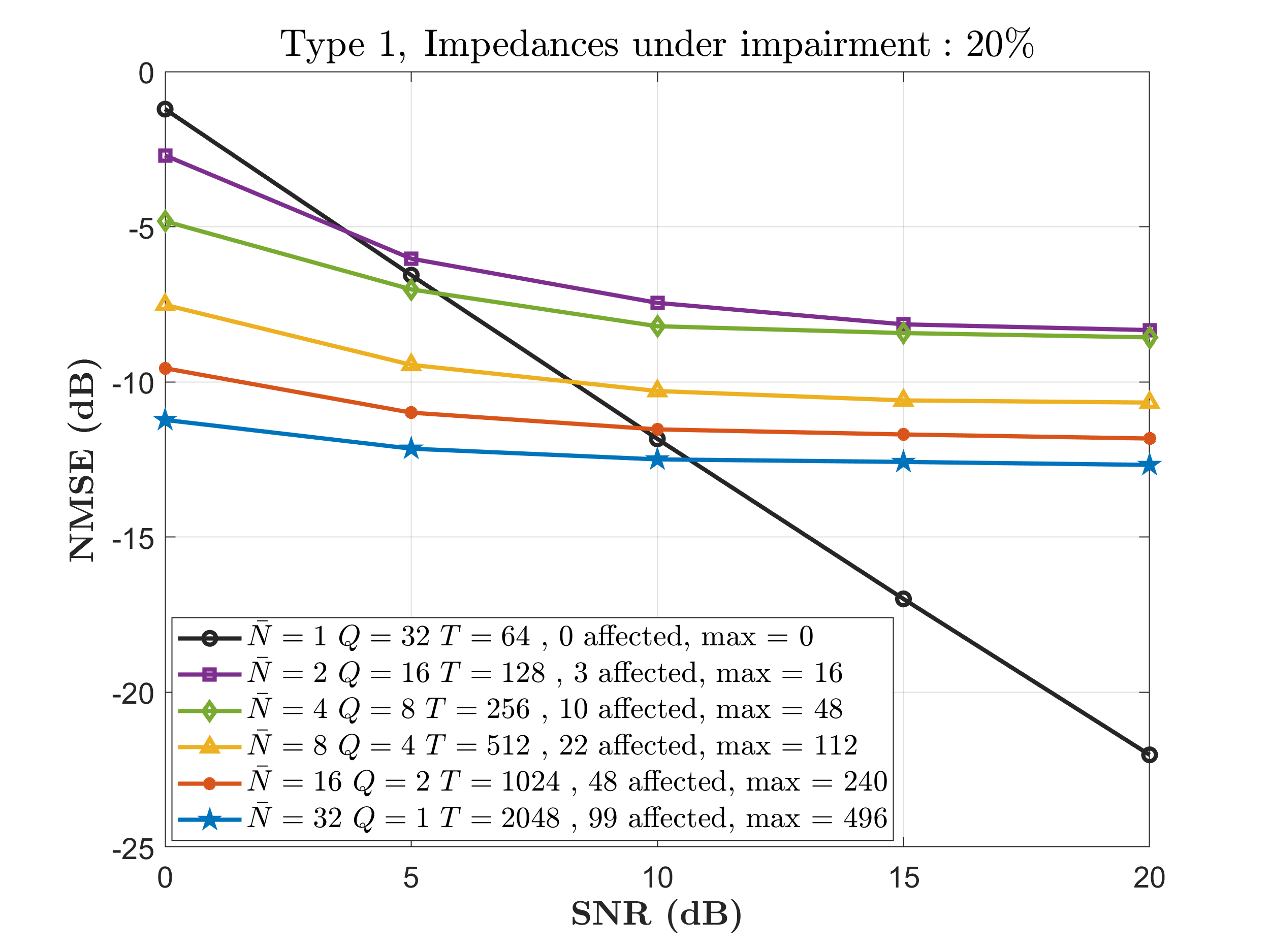}
    \caption{NMSE performance of BD-RIS under Type 1 impairment.}
    \label{fig:type1}
\end{figure}

\begin{figure}[!t]
    \centering
    \includegraphics[width=0.41\textwidth]{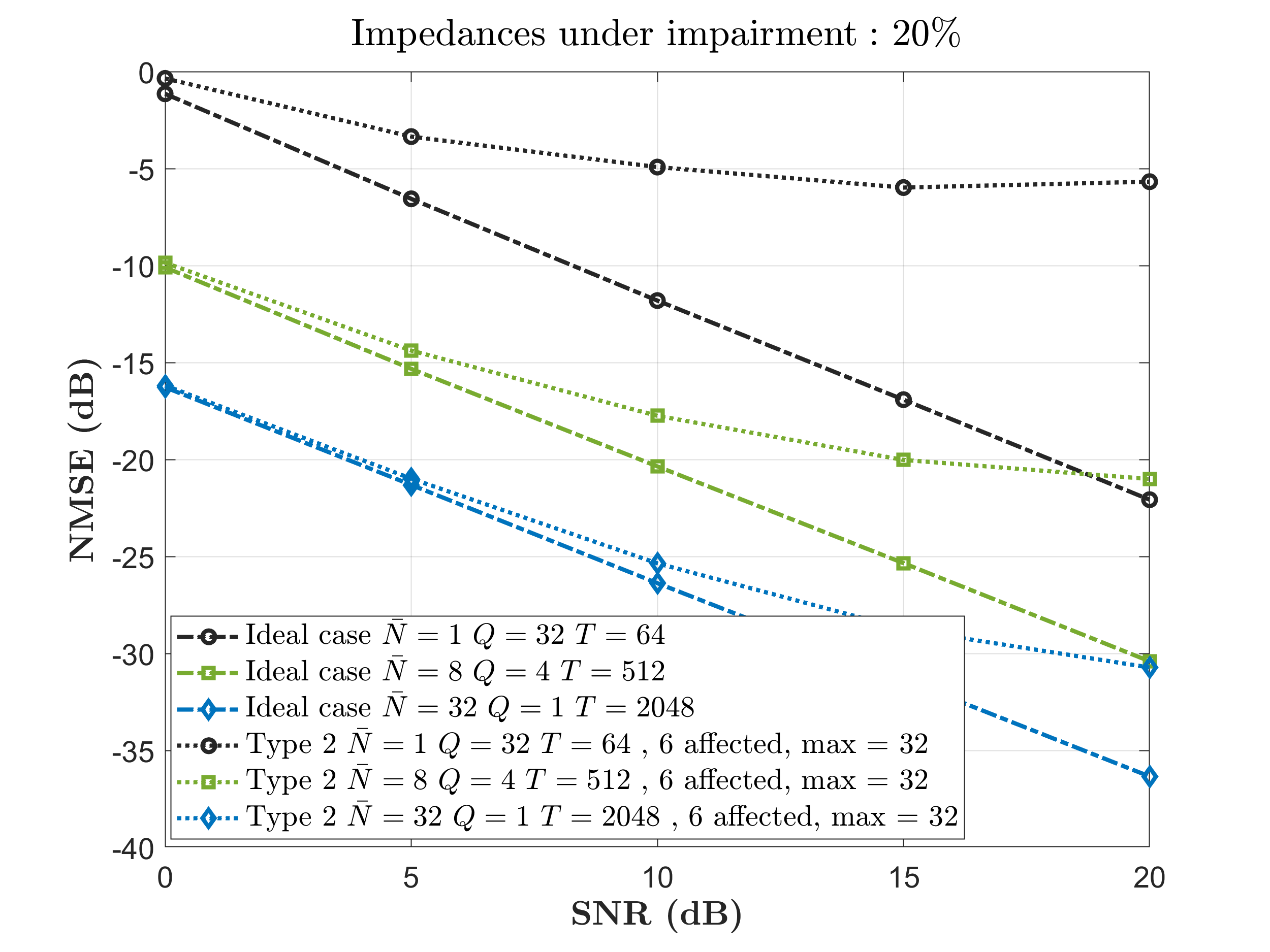}
    \caption{NMSE performance of ideal BD-RIS and Type 2 impairment.}
    \label{fig:ideal_and_type2}
\end{figure}

\begin{figure}[!t]
    \centering
    \includegraphics[width=0.41\textwidth]{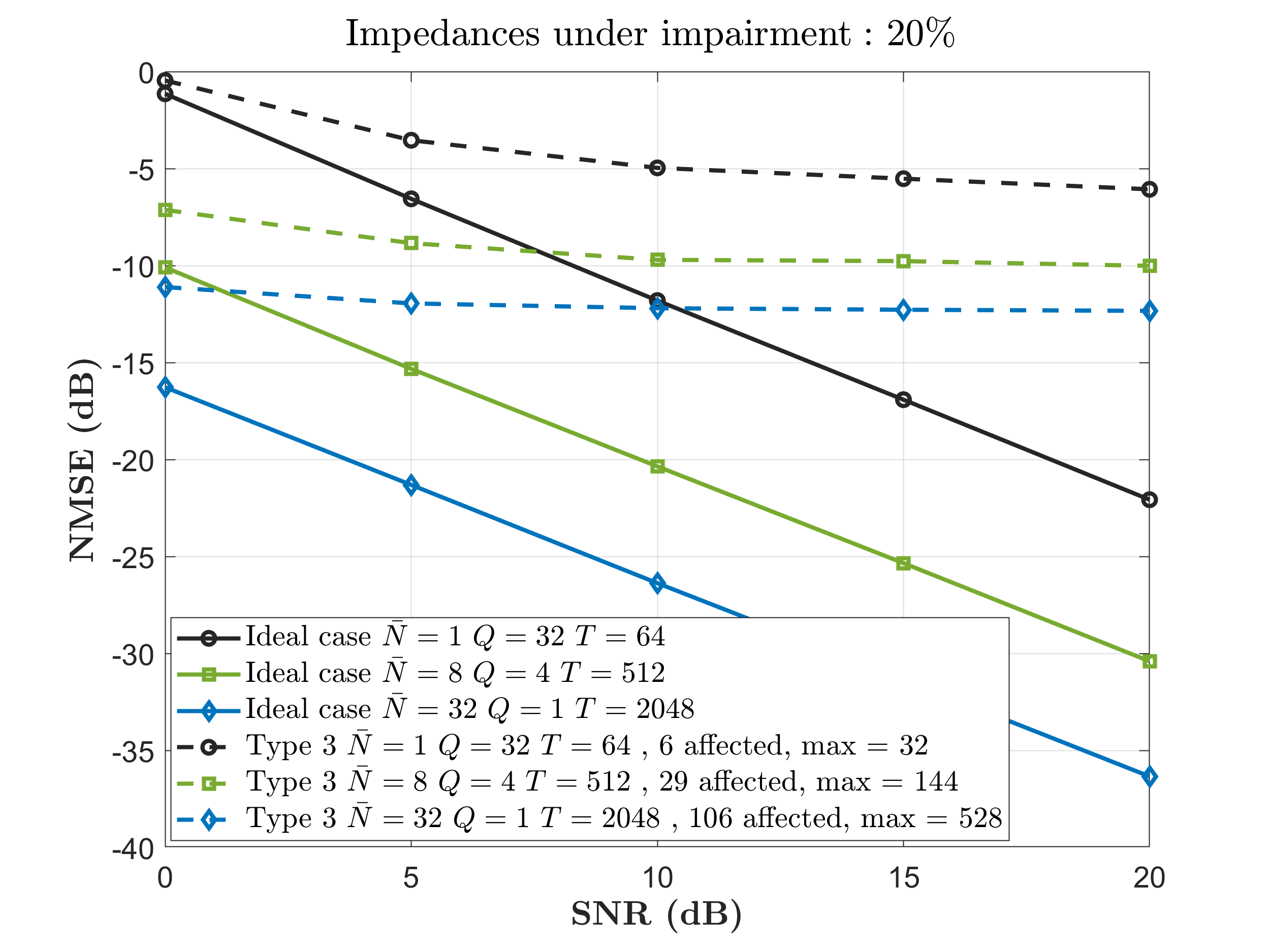}
    \caption{NMSE performance of ideal BD-RIS and Type 3 impairment.}
    \label{fig:ideal_and_type3}
\end{figure}

\begin{figure}[!t]
    \centering
    \includegraphics[width=0.41\textwidth]{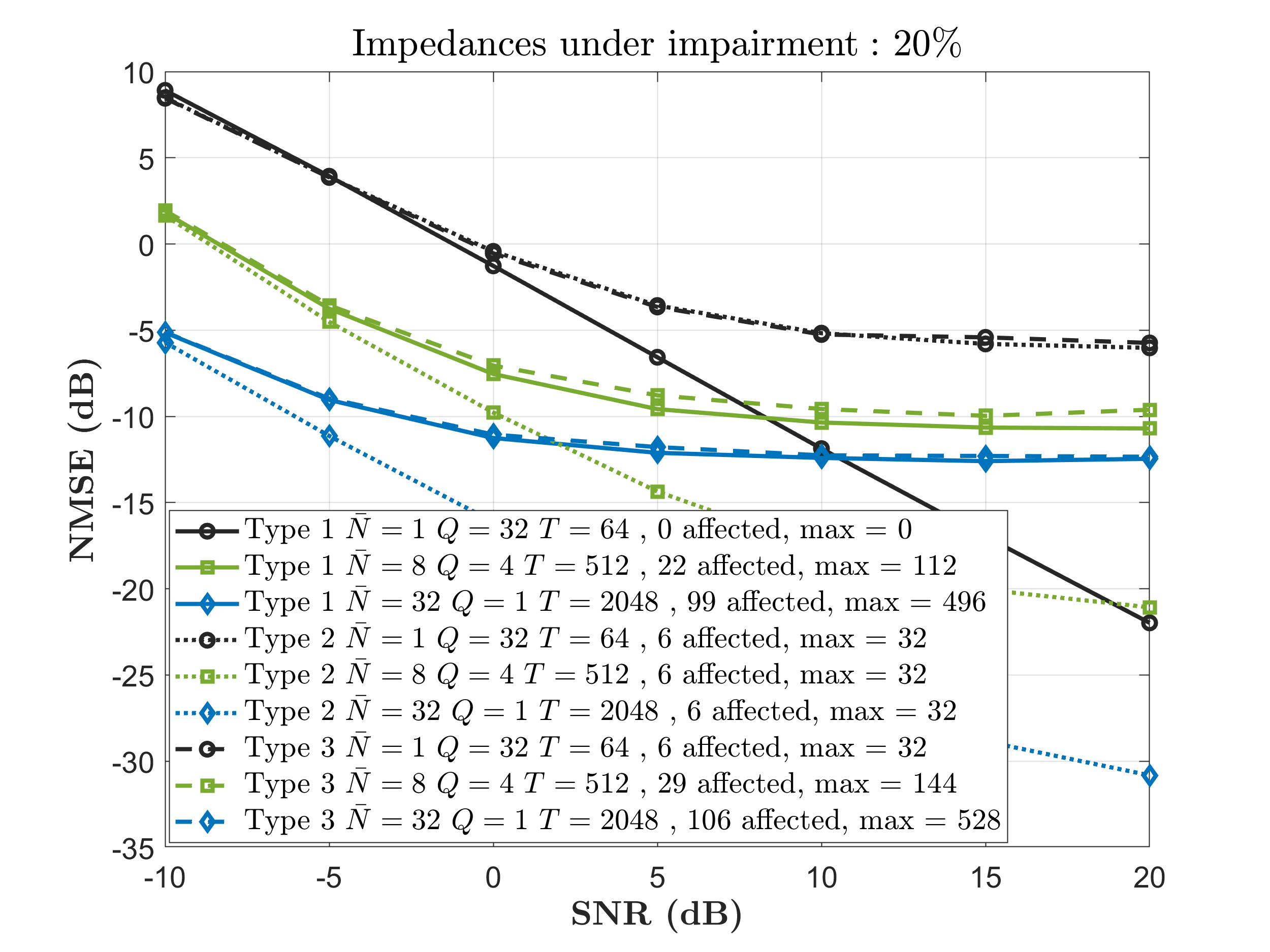}
    \caption{NMSE performance of BD-RIS under Type 1, 2, and 3 impairments.}
    \label{fig:types123}
\end{figure}

\begin{figure}[!t]
    \centering
    \includegraphics[width=0.41\textwidth]{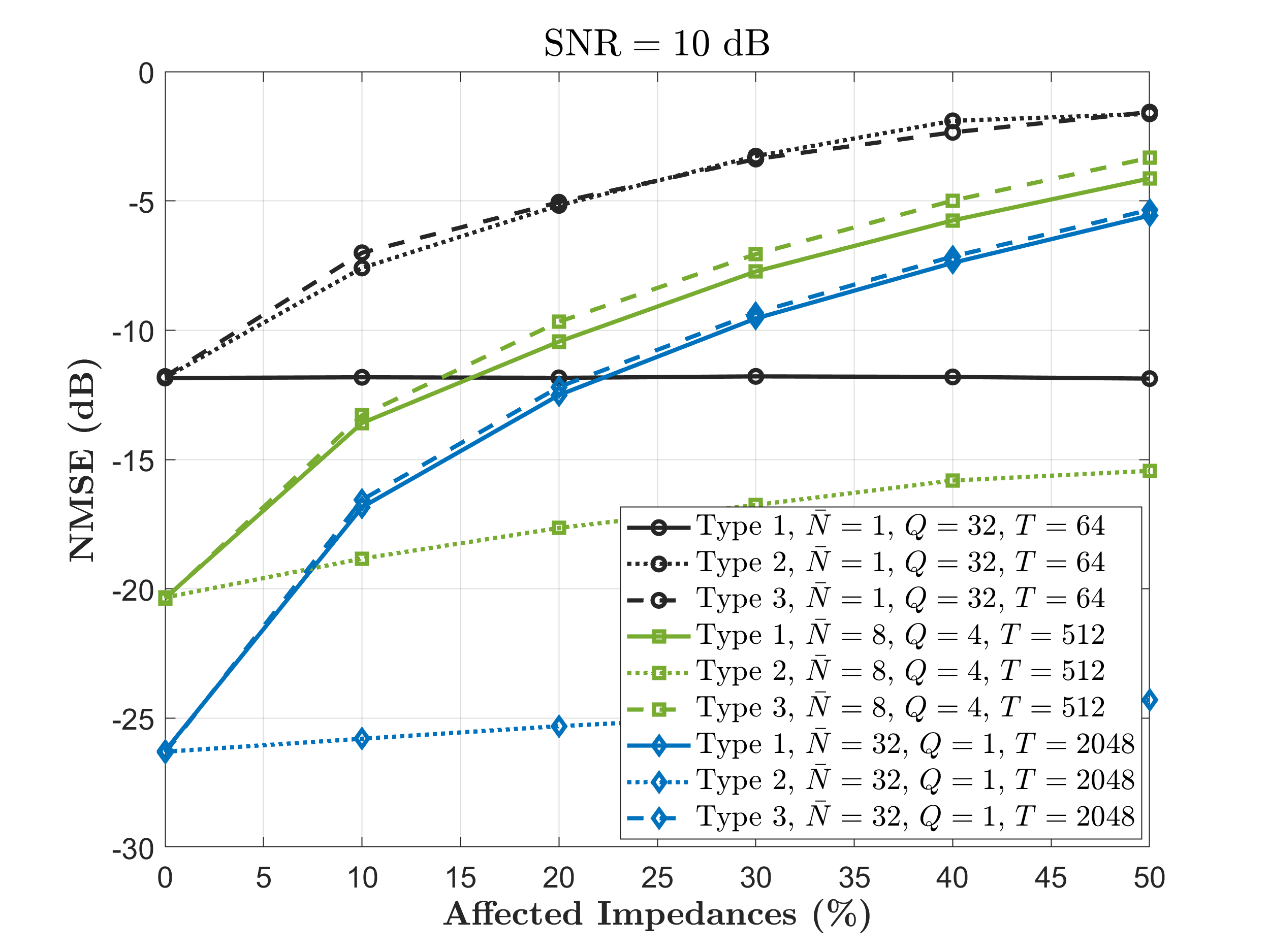}
    \caption{NMSE performance of BD-RIS under Type 1, 2, and 3 impairments for different percentages of affected impedances.}
    \label{fig:0a50type123}
\end{figure}

In this section, we evaluate the impact of the three proposed impairment models on the performance of the BD-RIS-assisted communication system. The analysis is based on Monte Carlo simulations over \(K=100\) independent trials. A new channel and a new impairment matrix are generated in each realization, allowing us to assess the average performance under different distortion conditions. The metric used for performance evaluation is the \ac{NMSE} of the estimated combined channel \(\ma{c} \in \mathbb{C}^{M_R M_T  \bar{N}^2 Q} \times 1\), defined as:
\[
\text{NMSE} = \frac{1}{K} \sum_{k=1}^{K} \frac{ \left\| \ma{c}_{(k)} - \hat{\ma{c}}_{(k)} \right\|_2^2 }{ \left\| \ma{c}_{(k)} \right\|_2^2 },
\]
where \(\hat{\ma{C}}_{(k)}\) denotes the  vectorized combined channel at the $k$-th trial. The BD-RIS training matrix \(\mathbf{S}\) is constructed using the orthogonal design proposed in \cite{bd_ris_group}. We assume orthogonal pilot sequences constructed from a truncated discrete Fourier transform (DFT) matrix \(\mathbf{X} \in \mathbb{C}^{M_T \times T}\), such that \(\mathbf{X}^{\mathrm{H}}\mathbf{X} = T\mathbf{I}_{M_T}\).  The pilot overhead is given by \(T = M_T \bar{N}^2 Q = M_T N \bar{N}\), where the group size \(\bar{N}\) plays a crucial role. As \(\bar{N}\) changes, the number of pilot symbols required also varies, even when the total number of \ac{RIS} elements \(N\) remains fixed. In all experiments, we assume an \ac{RIS} panel with \(N=32\) elements, \(M_T=2\)  and \(M_R=4\) antennas at the transmitter and receiver, respectively. The caption ‘max’  on Figs. \ref{fig:type1}-\ref{fig:0a50type123} represents the maximum number of impedances that can be affected, while the term ‘affected’ refers to the actual number of affected impedances on that experiment.

In Figure \ref{fig:type1}, we compare the channel estimation \ac{NMSE} performance by assuming a total of \(20\%\) impedances affected for the Type $1$ impairment. Since the total number of affected elements of $\ma{S}$ is given by $(\bar{N}^2Q - N)/2$, as the group size $\bar{N}$ increases, more disturbed elements are present in $\ma{S}$. We observe that the additive noise dominates the channel estimation performance for low \ac{SNR} values, between approximately $-10$dB and $3$dB. In other words, the effect of the hardware impairment is relatively minor compared to the noise power in this case. As a result, the \ac{BD-RIS} performance does not degrade significantly, and it can still outperform a conventional diagonal RIS (\(\bar{N}=1\)). However, as the \ac{SNR} increases, the influence of noise diminishes, and the effect of the impairment becomes dominant. This leads to an NMSE performance saturation in BD-RIS configurations with many affected off-diagonal elements. On the other hand, the conventional \ac{RIS} (when $\bar{N} =1$) is not affected by the impairment Type $1$ since this impairment, as shown in Figure~\ref{fig:impairments}, only affects off-diagonal elements.

In Figures \ref{fig:ideal_and_type2} and \ref{fig:ideal_and_type3}, we compare the \ac{NMSE} performance of the ideal RIS scenario, i.e., the case when the impairment matrix $\ma{E}$ is $\ma{1}_{N \times N}$, thus none element of $\ma{S}$ is affected, with the impairment Type $2$ (Fig. \ref{fig:ideal_and_type2}) and Type $3$ (Fig. \ref{fig:ideal_and_type3}), respectively. We assume that  \(20\%\) of the impedances are affected. As expected, in both cases, there is a consistent performance gap between the ideal and impaired cases in all configurations. However, in Figure \ref{fig:ideal_and_type2}, this gap decreases as \(\bar{N}\) increases. For \(\bar{N} = 1\), the \ac{NMSE} degradation is more present, while for \(\bar{N} = 32\), the performance loss is significantly reduced. This behavior comes from the fact that even when the same number of impedances are affected, the Type $2$ impairment only affects the diagonal elements of the scattering $\ma{S}$ (i.e., the self-impedances, see Fig. \ref{fig:impairments}). In contrast, the Type $3$ impairment affects all the elements of $\ma{S}$.

In Figure \ref{fig:types123}, we compare the three types of impairments by considering that  \(20\%\) of the BD-RIS impedances are affected. We can observe that, compared to the Type $1$ and Type $3$ impairments, the Type $2$ impairment less significantly affect the channel estimation accuracy. This is explained by the fact that the Type $2$ impairment only affects the diagonal elements of the scattering matrix. Thus, it is fixed over $N$ elements. In contrast, Type $1$ and Type $3$ affects up to $(\bar{N}^2Q - N)/2$ and $N + (\bar{N}^2Q - N)/2$, respectively.
% We see that Type 2 impairment, which affects only the diagonal elements, has the least negligible impact on the system performance when setting a fixed percentage of the maximum number of affected impedances. In contrast, Types 1 and 3 lead to much more pronounced degradation. Since Type 3 affects more matrix components, it performs slightly worse than Type 1.

In Figure \ref{fig:0a50type123}, we compare the impact of the three impairment models on the channel estimation accuracy by fixing the \ac{SNR} at $10$ dB and varying the number of affected impedances. As the percentage of affected impedances increases, all impairment types show performance degradation regarding channel estimation accuracy. However, Type $2$ impairment consistently leads to lower \ac{NMSE} values, confirming that it is the least harmful. In contrast, Type 1 and Type 3 produce a more significant degradation, with Type 3 performing slightly worse due to the larger number of affected matrix elements. This experiment highlights how different hardware impairments impact \ac{BD-RIS} performance depending on where the impairments occur in the scattering structure.

\section{Conclusion}
This work examined the practical impact of hardware imperfections on \acf{BD-RIS}. By introducing three impairment models on the ideal scattering matrix, we quantified their direct effect on channel estimation accuracy through the NMSE metric. Our results reveal that even a modest impairment level (e.g., \(20\%\) of affected impedances) can cause significant performance loss. This degradation becomes more pronounced as the \ac{SNR} increases, leading to a growing performance gap compared to the ideal case. The Type $1$ and Type $3$  impairments produce the most severe degradation due to the capacity of affecting more elements of the scattering matrix in a BD-RIS. As a perspective of this work, we shall investigate methods to estimate the impairment matrix $\ma{E}$ at the receiver. Identifying imperfect/malfunctioning elements is particularly useful in a real-world setup since it would allow us to mitigate their effect when designing the active and passive beamformings or to dynamically select/deactivate BD-RIS elements to optimize the system performance under practical imperfections.

% Moreover, we identified operating regimes where a conventional diagonal RIS outperforms the BD-RIS. These findings underscore the critical need for careful hardware design and impairment mitigation strategies when deploying BD‑RIS in real‑world MIMO systems.

% \section{Acknowledgements}
% The authors thank the partial support of FUNCAP under grant ITR-0214-00041.01.00/23, and the National Institute of Science and Technology (INCT-Signals) sponsored by Brazil's National Council for Scientific and Technological Development (CNPq) under grant 406517/2022-3. This work is also partially supported by CNPq under grant 312491/2020-4 and by Ericsson Research, Sweden, and Ericsson Innovation Center, Brazil, under UFC.52
% Technical Cooperation Contract Ericsson/UFC.

\bibliographystyle{IEEEtran}

\begin{thebibliography}{10}
\providecommand{\url}[1]{#1}
\csname url@samestyle\endcsname
\providecommand{\newblock}{\relax}
\providecommand{\bibinfo}[2]{#2}
\providecommand{\BIBentrySTDinterwordspacing}{\spaceskip=0pt\relax}
\providecommand{\BIBentryALTinterwordstretchfactor}{4}
\providecommand{\BIBentryALTinterwordspacing}{\spaceskip=\fontdimen2\font plus
\BIBentryALTinterwordstretchfactor\fontdimen3\font minus \fontdimen4\font\relax}
\providecommand{\BIBforeignlanguage}[2]{{%
\expandafter\ifx\csname l@#1\endcsname\relax
\typeout{** WARNING: IEEEtran.bst: No hyphenation pattern has been}%
\typeout{** loaded for the language `#1'. Using the pattern for}%
\typeout{** the default language instead.}%
\else
\language=\csname l@#1\endcsname
\fi
#2}}
\providecommand{\BIBdecl}{\relax}
\BIBdecl

\bibitem{basar_wireless_ris}
E.~Basar, M.~Di~Renzo, J.~De~Rosny, M.~Debbah, M.-S. Alouini, and R.~Zhang, ``Wireless communications through reconfigurable intelligent surfaces,'' \emph{IEEE access}, vol.~7, pp. 116\,753--116\,773, 2019.

\bibitem{gil_parafac}
G.~T. de~Araújo and A.~L.~F. de~Almeida, ``Parafac-based channel estimation for intelligent reflective surface assisted mimo system,'' in \emph{2020 IEEE 11th Sensor Array and Multichannel Signal Processing Workshop (SAM)}, 2020, pp. 1--5.

\bibitem{tensor_irs}
G.~T. de~Ara{\'u}jo, A.~L. de~Almeida, and R.~Boyer, ``Channel estimation for intelligent reflecting surface assisted mimo systems: A tensor modeling approach,'' \emph{IEEE Journal of Selected Topics in Signal Processing}, vol.~15, no.~3, pp. 789--802, 2021.

\bibitem{survey_irs}
B.~Zheng, C.~You, W.~Mei, and R.~Zhang, ``A survey on channel estimation and practical passive beamforming design for intelligent reflecting surface aided wireless communications,'' \emph{IEEE Communications Surveys \& Tutorials}, vol.~24, no.~2, pp. 1035--1071, 2022.

\bibitem{bd_ris_magazine}
H.~Li, S.~Shen, M.~Nerini, and B.~Clerckx, ``Reconfigurable intelligent surfaces 2.0: Beyond diagonal phase shift matrices,'' \emph{IEEE Communications Magazine}, vol.~62, no.~3, pp. 102--108, 2023.

\bibitem{shen_analysis}
S.~Shen, B.~Clerckx, and R.~Murch, ``Modeling and architecture design of reconfigurable intelligent surfaces using scattering parameter network analysis,'' \emph{IEEE Transactions on Wireless Communications}, vol.~21, no.~2, pp. 1229--1243, 2021.

\bibitem{bd_ris_architecture}
H.~Li, S.~Shen, and B.~Clerckx, ``Beyond diagonal reconfigurable intelligent surfaces: From transmitting and reflecting modes to single-~, group-~, and fully-connected architectures,'' \emph{IEEE Transactions on Wireless Communications}, vol.~22, no.~4, pp. 2311--2324, 2022.

\bibitem{li_dynamic_bd_ris}
{Li, Hongyu and Shen, Shanpu and Clerckx, Bruno}, ``A dynamic grouping strategy for beyond diagonal reconfigurable intelligent surfaces with hybrid transmitting and reflecting mode,'' \emph{IEEE Transactions on Vehicular Technology}, vol.~72, no.~12, pp. 16\,748--16\,753, 2023.

\bibitem{bd_ris_group}
H.~Li, Y.~Zhang, and B.~Clerckx, ``Channel estimation for beyond diagonal reconfigurable intelligent surfaces with group-connected architectures,'' in \emph{2023 IEEE 9th International Workshop on Computational Advances in Multi-Sensor Adaptive Processing (CAMSAP)}, 2023, pp. 21--25.

\bibitem{almeida_sokal_bd_ris_ce_TSP_2025}
A.~L.~F. de~Almeida, B.~Sokal, H.~Li, and B.~Clerckx, ``Channel estimation for beyond diagonal ris via tensor decomposition,'' \emph{IEEE Transactions on Signal Processing}, pp. 1--15, 2025.

\bibitem{li_channel_estimation}
H.~Li, S.~Shen, Y.~Zhang, and B.~Clerckx, ``Channel estimation and beamforming for beyond diagonal reconfigurable intelligent surfaces,'' \emph{IEEE Transactions on Signal Processing}, vol.~72, pp. 3318--3332, 2024.

\bibitem{decoupled_bd_ris}
B.~Sokal, Fazal-E-Asim, A.~L.~F. de~Almeida, H.~Li, and B.~Clerckx, ``A decoupled channel estimation method for beyond diagonal ris,'' in \emph{2024 58th Asilomar Conference on Signals, Systems, and Computers}, 2024, pp. 1395--1399.

\bibitem{varactor_model}
L.~C. d.~P. Pessoa, G.~T.~d. Araújo, P.~R.~B. Gomes, and A.~L. F.~d. Almeida, ``Irs-assisted communications under practical channel estimation and hardware model,'' in \emph{Anais do XXXIX Simpósio Brasileiro de Telecomunicações e Processamento de Sinais (SBrT)}, Fortaleza, CE, Brasil, setembro 2021, pp. 1--6.

\bibitem{paulo_impairments}
P.~R.~B. Gomes, G.~T. de~Araújo, B.~Sokal, A.~L. F.~d. Almeida, B.~Makki, and G.~Fodor, ``Channel estimation in ris-assisted mimo systems operating under imperfections,'' \emph{IEEE Transactions on Vehicular Technology}, vol.~72, no.~11, pp. 14\,200--14\,213, 2023.

\bibitem{imperfection_paulo}
P.~R.~B. Gomes, G.~T. de~Araújo, B.~Sokal, A.~L.~F. de~Almeida, B.~Makki, and G.~Fodor, ``Tensor-based channel estimation for ris-assisted communications with non-ideal phase shift responses,'' in \emph{2022 Workshop on Communication Networks and Power Systems (WCNPS)}, 2022, pp. 1--6.

\end{thebibliography}

% \appendix
% Insert informations about the apendix here

\end{document}